\documentstyle[epsfig]{aipproc}

\epsfclipon

\begin{document}
\title{Structure of the Goldstone Bosons}

\author{Roy J. Holt and Paul E. Reimer}
\address{Argonne National Laboratory\\
         Argonne, Illinois  60439}

\maketitle

\begin{abstract}
The feasibility of measuring the pion and kaon structure functions has
been investigated.  A high luminosity electron-proton collider would
make these measurements feasible.  Also, it appears feasible to
measure these structure functions in a nuclear medium.  Simulations
using the RAPGAP Monte Carlo of a possible pion structure function
measurement are presented.
\end{abstract}

Understanding hadron structure from the underlying quark and gluon
degrees of freedom and understanding modifications of hadrons in
nuclear matter are two of the most important goals of nuclear physics.
The light mesons have a central role in nucleon and nuclear structure.
The masses of the lightest hadrons, the mesons, are believed to arise
from explicit chiral symmetry breaking.  In particular, the light
mesons are the Goldstone bosons of quantum chromodynamics
\cite{mrt98}.  The pion, being the lightest meson, is particularly
interesting not only because of its importance in chiral perturbation
theory, but also because of its importance in explaining the quark sea
in the nucleon and the nuclear force in nuclei.

\section*{The Pion Structure Function}

\begin{figure}[tb] 
\centerline{\epsfig{file=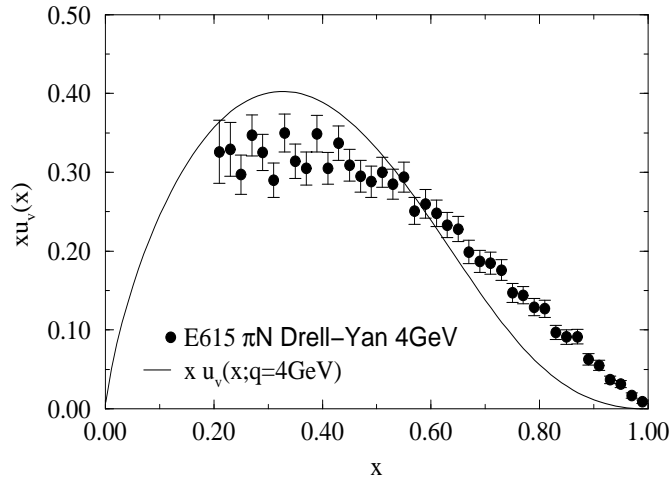,height=2.6in,width=3.5in}}
\caption{Existing data for the pion structure function from Drell-Yan
         scattering \protect\cite{conway}.  The solid curve represents
         a calculation of Hecht {\it et al.}~\protect\cite{hecht}. }
\label{fig1}
\end{figure}

At present, the pion is believed to contain a valence quark and
antiquark as well as a partonic sea.  Several theoretical calculations
are aimed at explaining the pion structure function in the valence
region.  These include Dyson-Schwinger\cite{hecht} and Nambu
Jona-Lasinio models\cite{shigetani,davidson}.  Lower order moments of
the structure function were determined in lattice gauge
calculations\cite{best}.  Typical agreement with the pion structure
function is shown in Fig.~\ref{fig1}.  Here, a curve from the
Dyson-Schwinger model is compared with the data from a pionic
Drell-Yan experiment\cite{conway} in the valence region.  The general
features of the valence structure of the pion are qualitatively
understood.  However, there is no good understanding of the pion sea.

Recently, measurements of the pion structure function\cite{adloff} at
very low $x$, in the region of the pion sea have been performed.  The
results of this work show two interesting findings: (1) the sea in the
pion has the same shape in $x$ as the sea in the proton, and (2) the
pion sea has approximately one-third of the magnitude as the sea in
the proton.  This latter result is especially surprising since one
expects that the pion sea to be two-thirds the value of the proton
sea.  These findings are even more surprising from the viewpoint of a
chiral quark model\cite{weise}.  This model predicts that the pion sea
carries a larger momentum fraction than the proton sea.  It appears
that a comparison of the sea in the pion and the proton is a clue to
understanding the nonperturbative structure of constituent quarks.
Thus, it is essential to measure the pion structure function,
especially the sea component, throughout the $x$ region from 0.02, the
highest value of the HERA data, up to 0.3, the lower value of the
Drell-Yan data.  This will map out the sea region where models should
have a high degree of validity.  Also, since there appears to be a
discrepancy between the data and the theoretical calculation at very
high $x$, another measurement using a different technique at high $x$
would be important.

\section*{Kaon Structure Function}

The valence structure of the kaon is comprised of a light $u$ or $d$
quark/antiquark and a strange quark/antiquark.  If our understanding
of the meson structure is correct, then the large difference between
the strange quark and $u$ or $d$ quark masses gives rise to a very
interesting effect for the kaon structure function.  In this case, the
strange quark, because of its large mass, carries more of the kaon's
momentum than the $u$ quark, say.  Then, the $u_v$ quark distribution
in the kaon should be shifted lower in $x$ than that in the pion.

\begin{figure} 
\centerline{\epsfig{file=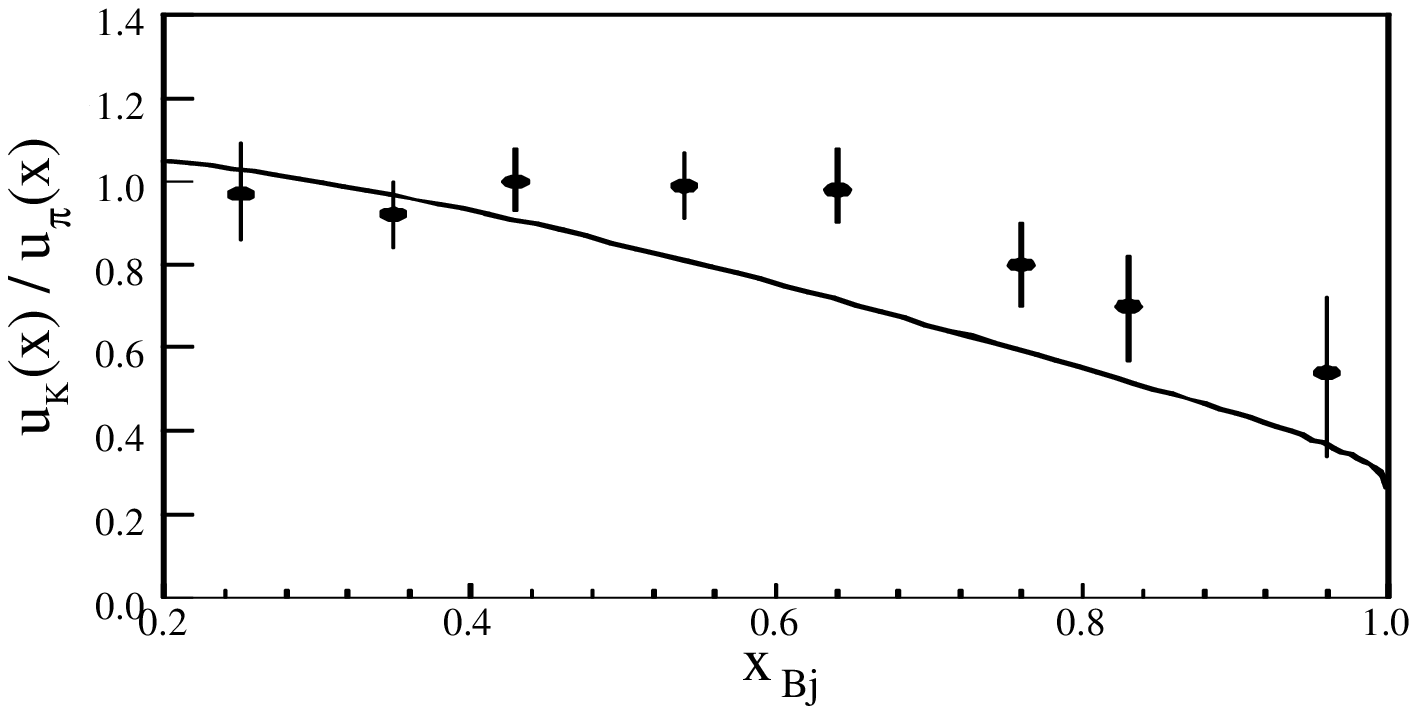, width=3.5in, height = 2.7 in}}
\caption{Existing data for $K^-/\pi^-$ ratio from Drell-Yan
         scattering \protect\cite{badier}.  The solid curve represents
         a calculation of Suzuki~\protect\cite{suzuki,shigetani}. }
\label{fig2}
\end{figure}

A Nambu Jona-Lasinio calculation\cite{shigetani,davidson} exhibits this
behavior as shown in Fig.~\ref{fig2}.  Here, the ratio of the valence
u quark in the kaon to that in the pion is shown as the curve in the
figure.  Drell-Yan measurements\cite{badier} of the ratio of K$^-$ to
$\pi^-$ shows a consistency with unity over most of the $x$ region, with
a suggestion that the ratio is dropping at high $x$.  However, the data
are not of sufficient quality to verify our understanding of this
process.  Thus, it is essential to measure the structure function at
high $x$ as well as in the sea region.

\section*{Meson Structure in the Nuclear Medium}

\begin{figure} 
\centerline{\epsfig{file=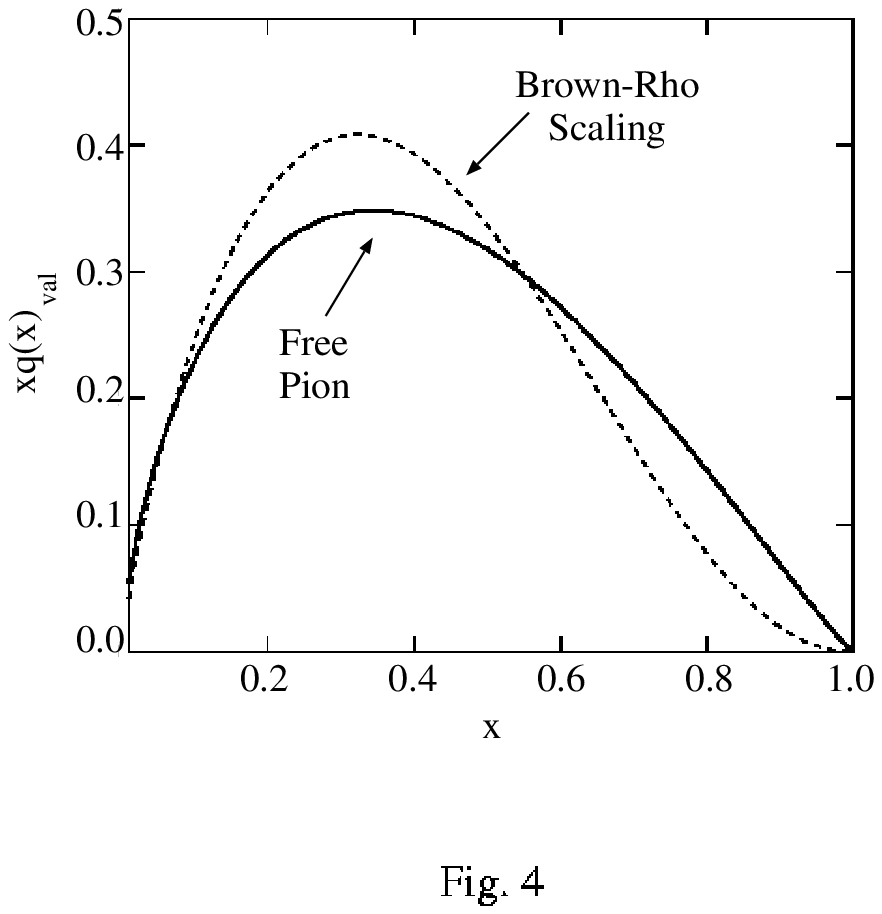, height=2.6 in, width=3.5 in}}

\caption{The solid curve represents the NJL calculation in
a nuclear medium, while the dashed curve gives the effect of 
Brown-Rho scaling in nuclear matter~\protect\cite{suzuki}.}

\label{fig3}
\end{figure}

The role of pions, and particularly, pion excess in the nuclear medium
have been long-standing issues in nuclear physics.  The pion excess
has not been observed in either Drell-Yan experiments \cite{drellyan}
at FNAL or in A(e,e'$\pi$) reactions\cite{jackson} at Jefferson Lab.
The main question is whether the pion or kaon structure function in a
nuclear medium is modified from the free structure function.
Calculations within the framework of a Nambu Jona-Lasinio
model\cite{suzuki} indicate that the medium modification for the pion
structure function should be small.  However, if one invokes Brown-Rho
scaling\cite{brown}, then the effect is large.  These calculations are
shown in Fig.~\ref{fig3}.  Here the NJL calculation is the solid
curve, while the dashed curve represents the Brown-Rho scaling.

\section*{Measurement of the Meson Structure Functions}

\begin{figure}[b] 
\centerline{\epsfig{file=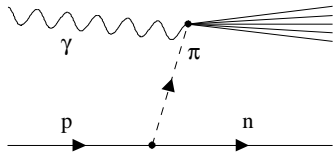}}
\vspace{10pt}

\caption{Deep inelastic scattering from the pion cloud surrounding a 
         proton.}

\label{fig4}
\end{figure}

The scattering process illustrated in Fig.~\ref{fig4} was simulated
using the RAPGAP Monte Carlo program~\cite{jung}.  RAPGAP models
processes in which there is a large rapidity gap between a fast
outgoing nucleon and the remainder of the inelastic scattering
fragments.  These processes include DIS from an exchanged
pion\cite{pirner,holtmann,kopeliovich} or pomeron..  A comparison of
results from RAPGAP with HERA data show reasonable agreement for fast
outgoing neutrons \cite{adloff}.

\begin{figure} 
\centerline{\epsfig{file=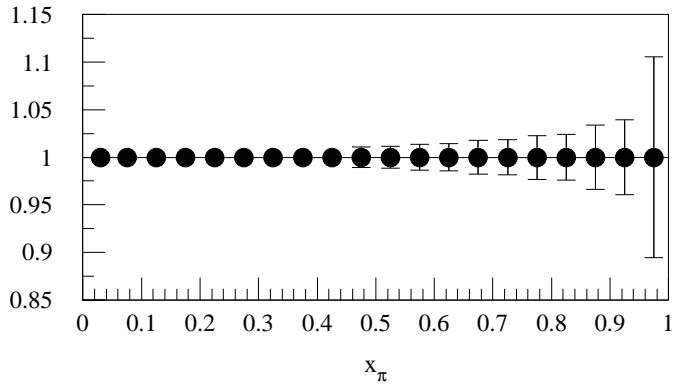}}
\caption{Simulated errors for DIS events using a 5 GeV electron beam
         on a 25 GeV proton beam with a luminosity of
         $10^{32}\textrm{cm}^{-2}\textrm{s}^{-1}$ and $10^6\textrm{s}$
         of running.}
\label{errors}
\end{figure}

For a 5 GeV electron beam on a 25 GeV proton beam, RAPGAP calculates
22 nb cross section for the $eq\rightarrow e'q'$ process.  The
expected accuracy of the experiments was calculated using events in
which a ``spectator'' neutron was identified.  In general, the neutron
is scattered less than 50 mrad from the nominal proton beam axis.
Events were cut on $-q^2 > 1~\textrm{GeV}^2$.  The expected errors are
shown in Fig.~\ref{errors}.  A luminosity of
$10^{32}~\textrm{cm}^{-2}\textrm{s}^{-1}$ was assumed for a run
lasting $10^6~\textrm{s}$.

The $K^+$ structure function can be measured by considering deep
inelastic scattering from the kaon cloud surrounding a proton.  The
basic Feynman diagram would be the same as in Fig.~\ref{fig4} with the
pion replaced by a kaon and the neutron replaced by a $\Lambda$.  The
probability for scattering from the $K^+$ cloud surrounding the proton
should be comparable to that for the $\pi^+$ since the KN$\Lambda$
coupling constant is comparable to that of the $\pi$NN vertex.  In
fact, one would only expect about a factor of two reduction in the
vertex function for the kaon compared to the pion.

The difficulty with this process is in the detection of the $\Lambda$.
The $\Lambda$ decays predominantly (64\%) to a proton and a $\pi^-$.
Thus, a special forward proton spectrometer as well as a forward pion
spectrometer would be necessary.  This should be feasible since the
ZEUS and H1 experiments at HERA have already successfully employed
forward proton spectrometers.  When designing the ring magnets, one
should take into account the possibility of detecting both positively
and negatively charged forward going hadrons.  Simulations for this
part of the experiment would be necessary to optimize the detection
efficiency.

A collider should also render these studies feasible for nuclei up to
$^4$He.  For the pion case, the idea would be to detect all of the
forward going nucleons from the deep inelastic scattering from the
pion.  In the case of a deuterium target, for example, one would
detect both forward going neutrons or forward going protons, depending
on whether the DIS occurred from the $\pi^+$ or the $\pi^-$,
respectively.

\section*{Summary}

In summary, a measurement of the pion and kaon structure functions
over a large $x_{\pi/K}$ region was shown to be feasible with an
electron proton collider, where the electron energy is 5 GeV, the
proton energy is 25 GeV, and the luminosity is 10$^{32}$
cm$^{-2}$s$^{-1}$.  A collider will also open up other very
interesting possibilities such as a measurement of the meson structure
function in the nuclear medium.

\section*{Acknowledgments}

We wish to thank C. Roberts, G. Levman, M. Derrick and T.-S. H. Lee
for very useful discussions.  In addition, we thank H. Jung and
D. H. Potterveld for valuable assistance with RAPGAP.  This work was
supported by the U. S. Department of Energy, Nuclear Physics Division,
under contract No. W-31-109-ENG-38.


\begin{references}

\bibitem{mrt98}Maris, P., Roberts, C.D., Tandy, P.C., {\it Phys. Lett.}
        {\bf B420}, 287 (1998).

\bibitem{hecht}Hecht, M. B., Roberts, C. D., Schmidt, S. M., preprint
        (2000), nucl-th/0008049.

\bibitem{shigetani}Shigetani, T., Suzuki, K., Toki, H. {\it Phys. Lett.}
         {\bf B308}, 383 (1993).

\bibitem{davidson}Davidson, R. M., Arriola, E. Ruiz
         {\it Phys. Lett} {\bf B348}, 163 (1995).

\bibitem{best}Best, C. {\it et al.} {\it Phys. Rev. } {\bf D56}, 2743
         (1997).

\bibitem{conway}Conway, J. S. {\it et al.} {\it Phys. Rev. } {\bf D39},
         39 (1989).

\bibitem{adloff}Adloff, C. {\it et al.} (H1 Collaboration)
        {\it Eur.  Phys. J. C } {\bf 6}, 587 (1999).

\bibitem{weise}Suzuki, K., Weise, W. {\it Nucl. Phys. } {\bf A634},
         141 (1998).

\bibitem{badier}Badier, J., {\it et al.} {\it Phys. Lett. } {\bf B93},
         354 (1980).

\bibitem{drellyan} Alde, D. M. {\it et al.} {\it Phys. Rev. Lett.} 
        {\bf 64}, 2479 (1990).

\bibitem{jackson} Jackson, H. E. (NucPi Collaboration) {\it Sixteenth 
        Int'l Conf. on Few Body Systems } Taipei, preprint (2000).

\bibitem{suzuki}Suzuki, K. {\it Phys. Lett. } {\bf B368}, 1 (1996).

\bibitem{brown}Brown, G. E., Rho, M. {\it Phys. Rev. Lett. } {\bf 66},
         2720 (1991).

\bibitem{jung} Jung, H., {\it Comp. Phys. Commun.} {\bf 86}, 147 (1995).

\bibitem{pirner}D'Alesio, U. and Pirner, H. J. {\it Eur. Phys. J. A }
         {\bf 7}, 109 (2000).

\bibitem{holtmann} Holtmann, H. {\it et al.} {\it Nucl. Phys.}
         {\bf A569}, 631 (1996).

\bibitem{kopeliovich} Kopeliovich, H. {\it et al.} {\it Z. Phys.  }
         {\bf 6}, 587 (1999).




\end{references}
\end{document}